\documentclass[11pt]{article}
\usepackage[totalwidth=465pt,totalheight=645pt]{geometry}

\usepackage{amsmath,amssymb,amsthm,bbm,graphicx,natbib}
\bibpunct{(}{)}{,}{a}{,}{,}
\newtheorem{theorem}{Theorem}
\newtheorem{assumption}[theorem]{Assumption}
\newtheorem{algorithm}[theorem]{Algorithm}
\newtheorem{example}[theorem]{Example}
\newcommand{\1}{{\mathbbm{1}}}
\newcommand{\E}{{\operatorname{E}}}
\renewcommand{\P}{{\operatorname{P}}}

\newcommand{\gaus}{{\operatorname{\mathcal{N}}}}

\newcommand{\ue}{{\mathrm{e}}}
\newcommand{\ud}{{\operatorname{d}}}

\title{Online Expectation-Maximisation\footnote{To appear in \emph{Mixtures}, edited by Kerrie Mengersen, Mike Titterington and Christian P. Robert.}}
\author{Olivier Capp\'{e}\\CNRS \& Telecom ParisTech, Paris, France}
\date{}

\begin{document}

\maketitle

\bigskip

\tableofcontents

\bigskip

Before entering into any more details about the methodological aspects, let's discuss the
motivations behind the association of the two phrases ``online (estimation)'' and
``Expectation-Maximisation (algorithm)'' as well as their pertinence in the context of mixtures and
more general models involving latent variables.

The adjective \emph{online} refers to the idea of computing estimates of model parameters
on-the-fly, without storing the data and by continuously updating the estimates as more
observations become available. In the machine learning literature, the phrase \emph{online
  learning} has been mostly used recently to refer to a specific way of analysing the performance
of algorithms that incorporate observations progressively \citep{cesabianchi_lugosi_06}. We dot not
refer here to this approach and will only consider the more traditional setup in which the
objective is to estimate fixed parameters of a statistical model and the performance is quantified
by the proximity between the estimates and the parameter to be estimated. In signal processing and
control, the sort of algorithms considered in the following is often referred to as \emph{adaptive}
or \emph{recursive} \citep{ljung_soderstrom_83,benveniste:metivier:priouret:1990}. The word
recursive is so ubiquitous in computer science that its use may be somewhat ambiguous and is not
recommended. The term adaptive may refer to the type of algorithms considered in this chapter but
is also often used in contexts where the focus is on the ability to track slow drifts or abrupt
changes in the model parameters, which will not be our primary concerns.

Traditional applications of online algorithms involve situations in which the data cannot be stored,
due to its volume and rate of sampling as in real-time signal processing or stream mining. The wide
availability of very large datasets involving thousands or millions of examples is also at the
origin of the current renewed interest in online algorithms. In this context, online algorithms are
often more efficient ---i.e., converging faster towards the target parameter value--- and need fewer
computer resource, in terms of memory or disk access, than their batch counterparts
\citep{neal:hinton:1999}. In this chapter, we are interested in both contexts: when the online
algorithm is used to process on-the-fly a potentially unlimited amount of data or when it is
applied to a fixed but large dataset. We will refer to the latter context as the \emph{batch
  estimation} mode.

Our main interest is maximum likelihood estimation and although we may consider adding a penalty
term (i.e., Maximum A Posteriori estimation), we will not consider ``fully Bayesian'' methods which
aim at sequentially simulating from the parameter posterior. The main motivation for this
restriction is to stick to computationally simple iterations which is an essential requirement of
successful online methods. In particular, when online algorithms are used for batch estimation, it
is required that each parameter update can be carried out very efficiently for the method to be
computationally competitive with traditional batch estimation algorithms. Fully Bayesian approaches
---see, e.g., \citep{chopin_02}--- typically require Monte Carlo simulations even in simple models
and raise some challenging stability issues when used on very long data records
\citep{kantas_doucet_singh_maciejowski_09}.

This quest for simplicity of each of the parameter update is also the reason for focussing on the
EM (Expectation-Maximisation) algorithm. Ever since its introduction by
\cite{dempster:laird:rubin:1977}, the EM algorithm has been criticised for its often sub-optimal
convergence behaviour and many variants have been proposed by, among
others, \cite{lange:1995,meng:vandyk:1997}. This being said, thirty years after the seminal paper
by Dempster and his coauthors, the EM algorithm still is, by far, the most widely used inference
tool for latent variable models due to its numerical stability and ease of implementation. Our main
point here is not to argue that the EM algorithm is always preferable to other options. But the EM
algorithm which does not rely on fine numerical tunings involving, for instance, line-searches,
re-projections or pre-conditioning is a perfect candidate for developing online versions with very
simple updates. We hope to convince the reader in the rest of this chapter that the online version
of EM that is described here shares many of the attractive properties of the original proposal of
\cite{dempster:laird:rubin:1977} and provides an easy to implement and robust solution for online
estimation in latent variable models.

Quite obviously, guaranteeing the strict likelihood ascent property of the original EM algorithm is
hardly feasible in an online context. On the other hand, a remarkable property of the online EM
algorithm is that it can reach asymptotic Fisher efficiency by converging towards the actual
parameter value at a rate which is equivalent to that of the Maximum Likelihood Estimator
(MLE). Hence, when the number of observations is sufficiently large, the online EM algorithm does
becomes highly competitive and it not necessary to consider potentially faster-converging
alternatives. When used for batch estimation, i.e., on a fixed large dataset, the situation is more
contrasted but we will nonetheless show that the online algorithm does converge towards the maximum
likelihood parameter estimate corresponding to the whole data. To achieve this result, one
typically needs to scan the data set repeatedly. In terms of computing time, the advantages of
using an online algorithm in this situation will typically depend on the size of the problem. For
long data records however, this approach is certainly more recommendable than the use of the
traditional batch EM algorithm and preferable to other alternatives considered in the literature.

The rest of this chapter is organised as follows. The first section is devoted to the modelling
assumptions that are adopted throughout the chapter. In Section~\ref{sec:limitingEM}, we consider the
large-sample behaviour of the traditional EM algorithm, insisting on the concept of the limiting EM
recursion which is instrumental in the design of the online algorithm. The various aspects of the
online EM algorithm are then examined in Sections~\ref{sec:onlineEM} and~\ref{sec:disc}.

Although the chapter is intended to be self-contained, we will nonetheless assume that the reader is familiar with the fundamental concepts of classical statistical inference and, in particular, with Fisher information, exponential families of distributions and maximum likelihood estimation, at the level of \cite{lehmann:casella:1998,bickel:docksum:1977} or equivalent texts.

\section{Model and Assumptions}

We assume that we are able to observe an independent sequence of
identically distributed data $(Y_t)_{t\geq 1}$, with marginal distribution
$\pi$. An important remark is that we do not necessarily assume that $\pi$
corresponds to a distribution that is reachable by the statistical model that
is used to fit the data. As discussed in Section~\ref{sec:batch_ML} below,
this distinction is important to analyse the use of
the online algorithm for batch maximum-likelihood estimation.

The statistical models that we consider are of the missing-data type, with an
unobservable random variable $X_t$ associated to each observation $Y_t$. The
latent variable $X_t$ may be continuous or vector-valued and we will
not be restricting ourselves to finite mixture models. Following the terminology
introduced by \cite{dempster:laird:rubin:1977}, we will refer to the pair
$(X_t,Y_t)$ as the \emph{complete data}. The likelihood function
$f_{\theta_\star}$ is thus defined as the marginal
\[
  f_{\theta}(y_t) = \int p_\theta(x_t,y_t) d x_t \, ,
\]
where $\theta \in \Theta$ is the parameter of interest to be estimated. If the
actual marginal distribution of the data belongs to the family of the model
distributions, i.e., if $\pi = f_{\theta_\star}$ for some parameter value
$\theta_\star$, the model is said to be \emph{well-specified}; but, as
mentioned above, we dot not restrict ourselves to this case. In the following, 
the statistical model $(p_\theta)_{\theta\in\Theta}$ is assumed to verify the
following key requirements.

\begin{assumption}{Modelling Assumptions}
  \label{asum:main}
  \begin{enumerate}
  \item[(i)] The model belongs to a (curved) exponential family
  \begin{equation}
  \label{eq:complete_data_exp_family}
  p_\theta(x_t,y_t) = h(x_t,y_t) \exp\left(\langle s(x_t,y_t), \psi(\theta) \rangle - A(\theta)\right) \, ,
  \end{equation}
  where $s(x_t,y_t)$ is a vector of \emph{complete-data sufficient statistics} belonging to a convex set $\mathcal{S}$,
  $\langle \cdot, \cdot \rangle$ denotes the dot product and $A$ is the
  \emph{log-partition function}.
  \item[(ii)] The complete-data maximum-likelihood is explicit, in the sense that the function $\bar{\theta}$ defined by
  \begin{align*}
  & \bar{\theta} : \mathcal{S} \to \Theta \\
  & S \mapsto \bar{\theta}(S) = \arg\max_{\theta\in\Theta} \, \langle S, \psi(\theta) \rangle - A(\theta)
  \end{align*}
  is available in closed-form.
  \end{enumerate}
\end{assumption}

Assumption~\ref{asum:main} defines the context where the EM algorithm may be used directly (see, in
particular, the discussion of \citealp{dempster:laird:rubin:1977}). Note that
\eqref{eq:complete_data_exp_family} is not restricted to the specific case of exponential family
distributions in \emph{canonical} (or \emph{natural}) parameterisation. The latter correspond to
the situation where $\psi$ is the identity function, which is particular in that $p_\theta$ is then
log-concave with the complete-data Fisher information matrix $I_p(\theta)$ being given by the
Hessian $\nabla^2_\theta A(\theta)$ of the log-partition function. Of course, if $\psi$ is an
invertible function, one could use the reparameterisation $\eta = \psi(\theta)$ to recover the
natural parameterisation but it is important to recognise that for many models of interest, $\psi$
is a function that maps low-dimensional parameters to higher-dimensional statistics. To illustrate
this situation, we will use the following simple running example.

\begin{example}[Probabilistic PCA Model]
  \label{ex:pca1}
  Consider the \emph{probabilistic Principal Component Analysis (PCA)} model of
  \cite{tipping_bishop_99}. The model postulates that a $d$-dimensional observation vector $Y_t$
  can be represented as
  \begin{equation}
    \label{eq:prob_pca}
    Y_t = u X_t + \lambda^{1/2} N_t \, ,
  \end{equation}
  where $N_t$ is a centred unit-covariance $d$-dimensional multivariate Gaussian vector, while the
  latent variable $X_t$ also is such a vector but of much lower-dimension. Hence, $u$ is a $d
  \times r$ matrix with $r \ll d$.

  Eq.~\eqref{eq:prob_pca} is thus fully equivalent to assuming that $Y_t$ is a centred
  $d$-dimensional Gaussian variable with a structured covariance matrix given by $\Sigma(\theta) =
  uu'+ \lambda I_d$, where the prime denotes transposition and $I_d$ is the $d$-dimensional
  identity matrix. Clearly, there are in this model many ways to estimate $u$ and $\lambda$ that do
  not rely on the probabilistic model of~\eqref{eq:prob_pca}; the standard PCA being probably the
  most well-known. \cite{tipping_bishop_99} discuss several reasons for using the probabilistic
  approach that include the use of priors on the parameters, the ability to deal with missing or
  censored coordinates of the observations but also the access to quantitative diagnostics based on
  the likelihood that are helpful, in particular, for determining the number of relevant
  factors.

  To cast the model of~\eqref{eq:prob_pca} in the form given
  by~\eqref{eq:complete_data_exp_family}, the complete-data model
  \begin{equation*}
    \begin{pmatrix}
      X_t \\ \hdots \\ Y_t
    \end{pmatrix} \sim \gaus\left(\begin{pmatrix} 0 \\ \hdots \\ 0
      \end{pmatrix}, \begin{pmatrix}
        I_r  & \vdots & u' \\
        \hdotsfor{3}\\
        u & \vdots & uu' + \lambda I_d
      \end{pmatrix}\right)
  \end{equation*}
  must be reparameterised by the precision matrix $\Sigma^{-1}(\theta)$, yielding
  \[
  p_\theta(x_t,y_t) = (2\pi)^{-d/2} \exp \left[\operatorname{trace}\left\{\Sigma^{-1}(\theta)
      s(x_t,y_t)\right\} -\frac12 \log |\Sigma(\theta)|\right] \, ,
  \]
  where $s(x_t,y_t)$ is the rank one matrix
  \[
  s(x_t,y_t) = -\frac12 \begin{pmatrix} X_t \\ \hdots \\ Y_t
  \end{pmatrix} \begin{pmatrix} X_t \\ \hdots \\ Y_t
  \end{pmatrix}' \, .
  \]
  Hence in the probabilistic PCA model, the $\psi$ function in this case maps the pair
  $(u,\lambda)$ to the $(d+1)$-dimensional symmetric positive definite matrix
  $\Sigma^{-1}(\theta)$. Yet, Assumption~\ref{asum:main}-(ii) holds in this
  case and the EM algorithm can be used ---see the general formulas given in
  \citep{tipping_bishop_99} as well as the details of the particular case where $r=1$ below.

  In the following, we will more specifically look at the particular case where $r=1$ and $u$ is
  thus a $d$-dimensional vector. This very simple case also provides a nice and concise
  illustration of more complex situations, such as the \emph{Direction Of Arrival (DOA)} model
  considered by \cite{cappe:charbit:moulines:2006}. The case of of a single factor PCA is also
  interesting as most required calculations can be done explicitly. In particular, it is possible
  to provide the following expression for $\Sigma^{-1}(\theta)$ using the block matrix inversion
  and Sherman-Morrison formulas:
  \[
  \Sigma^{-1}(\theta) = \begin{pmatrix}
    1 + \|u\|^2/\lambda & \vdots & -u'/\lambda \\
    \hdotsfor{3}\\
    -u/\lambda & \vdots & \lambda^{-1} I_d
  \end{pmatrix} \, .
  \]
  The above expression shows that one may redefine the sufficient statistics $s(x_t,y_t)$ as
  consisting solely of the scalars $s^{(0)}(y_t) = \|y_t\|^2$, $s^{(2)}(x_t) = x_t^2$ and of the
  $d$-dimensional vector $s^{(1)}(x_t,y_t) = x_t y_t$. The complete-data log-likelihood is then given by
  \begin{equation}
   \label{eq:pca1_completelogl}
    \log p_\theta(x_t,y_t) = C^{st} -\frac12 \left\{ n d \log \lambda + \frac{s^{(0)}(y_t)-2 u' s^{(1)}(x_t,y_t)+s^{(2)}(x_t) \|u\|^2}{\lambda}\right\} \, ,
  \end{equation}
  ignoring constant terms that do not depend on the parameters $u$ and $\lambda$.

  Note that developing an online algorithm for~\eqref{eq:prob_pca} in this particular case is
  equivalent to recursively estimating the largest eigenvalue of a covariance matrix from a series
  of multivariate Gaussian observations. We return to this example shortly below.
\end{example}

\section{The EM Algorithm and the Limiting EM Recursion}
\label{sec:limitingEM}
In this section, we first review core elements regarding the EM algorithm that will be needed in
the following. Next, we introduce the key concept of the \emph{limiting EM recursion}
which corresponds to the limiting deterministic iterative algorithm obtained when the number of
observations grows to infinity. This limiting EM recursion is important both to understand the
behaviour of classic batch EM when used with many observations and for motivating the form of the online
EM algorithm.

\subsection{The Batch EM Algorithm}
\label{sec:batch_EM}
In light of Assumption~\ref{asum:main} and in particular of the assumed form of the likelihood in~(\ref{eq:complete_data_exp_family}), the classic EM algorithm of \cite{dempster:laird:rubin:1977} takes the following form.

\begin{algorithm}[Batch EM Algorithm]
\label{alg:batch_EM}
Given $n$ observations, $Y_1, \dots, Y_n$ and an initial parameter guess $\theta_0$, do, for $k \geq 1$,
\begin{description}
\item[E-Step]
  \begin{equation}
  S_{n,k} = \frac1n \sum_{t=1}^n \E_{\theta_{k-1}} \left[\left. s(X_t,Y_t) \right| Y_t\right] \, ,
  \label{eq:batchEM_E}
  \end{equation}
\item[M-Step]
  \begin{equation}
  \theta_{k} = \bar \theta\left(S_{n,k}\right) \, .
  \label{eq:batchEM_M}
  \end{equation}
\end{description}
\end{algorithm}

Returning to the single factor PCA model of Example~\ref{ex:pca1}, it is easy to check from the
expression of the complete-data log-likelihood in~\eqref{eq:pca1_completelogl} and the definition
of the sufficient statistics that the E-step reduces to the computation of
\begin{align}
  & s^{(0)}(Y_t) = \|Y_t\|^2 \, , \nonumber\\
  & \E_{\theta} \left[\left. s^{(1)}(X_t,Y_t) \right| Y_t\right] = \frac{Y_t Y_t' u}{\lambda + \|u\|^2} \, , \nonumber\\
  & \E_{\theta} \left[\left. s^{(2)}(X_t) \right| Y_t\right] = \frac{\lambda}{\lambda + \|u\|^2} + \frac{(Y_t' u)^2}{(\lambda + \|u\|^2)^2} \, , \label{eq:pca1_E}
\end{align}
with the corresponding empirical averages
\begin{align*}
  & S_{n,k}^{(0)} = \frac1n \sum_{t=1}^n s^{(0)}(Y_t) \, , \\
  & S_{n,k}^{(1)} = \frac1n \sum_{t=1}^n  \E_{\theta_{k-1}} \left[\left. s^{(1)}(X_t,Y_t) \right| Y_t\right] \, , \\
  & S_{n,k}^{(2)} = \frac1n \sum_{t=1}^n \E_{\theta_{k-1}} \left[\left. s^{(2)}(X_t) \right| Y_t\right] \, .   
\end{align*}
The M-step equations which define the function $\bar{\theta}$ are given by
\begin{align}
  & u_k = \bar{\theta}^{(u)}(S_{n,k}) = S_{n,k}^{(1)}/S_{n,k}^{(2)} \, , \nonumber \\
  & \lambda_k = \bar{\theta}^{(\lambda)}(S_{n,k}) = \frac1d \left\{S_{n,k}^{(0)} - \|S_{n,k}^{(1)}\|^2/S_{n,k}^{(2)}\right\} \, . \label{eq:pca1_M}
\end{align}

\subsection{The Limiting EM Recursion}
Returning to the general case, a very important remark is that Algorithm~\ref{alg:batch_EM} can
be fully reparameterised in the domain of sufficient statistics, reducing to the recursion
\[
S_{n,k} = \frac1n \sum_{t=1}^n \E_{\bar{\theta}(S_{n,k-1})} \left[\left. s(X_t,Y_t) \right|
  Y_t\right]
\]
with the convention that $S_{n,0}$ is such that $\theta_0 = \bar{\theta}(S_{n,0})$. Clearly, if an
uniform (wrt. $S \in \mathcal{S}$) law of large numbers holds for the empirical averages of
$\E_{\bar{\theta}(S)} \left[\left. s(X_t,Y_t) \right| Y_t\right]$, the EM update tends, as the
number of observations $n$ tends to infinity, to the deterministic mapping $T_{\mathcal{S}}$ on
$\mathcal{S}$ defined by
\begin{equation}
  \label{eq:limitingEM}
  T_{\mathcal{S}}(S) = \E_\pi \left(\E_{\bar{\theta}(S)} \left[\left. s(X_1,Y_1) \right| Y_1\right] \right) \, .  
\end{equation}
Hence, the sequence of EM iterates $(S_{n,k})_{k\geq 1}$ converges to the sequence $(T^k_{\mathcal{S}}(S_0))_{k\geq 1}$, which is deterministic except for the choice of $S_0$. We refer to the limiting mapping $T_{\mathcal{S}}$ defined in~\eqref{eq:limitingEM}, as the \emph{limiting EM recursion}. Of course, this mapping on $\mathcal{S}$ also induces a mapping $T_{\Theta}$ on $\Theta$ by considering the values of $\theta$ associated to values of $S$ by the function $\bar{\theta}$. This second mapping is defined as
\begin{equation}
  \label{eq:limitingEM_param}
  T_{\Theta}(\theta) = \bar \theta\left\{ \E_\pi \left(\E_{\theta} \left[\left. s(X_1,Y_1) \right|
        Y_1\right] \right) \right\} \, .
\end{equation}
Using exactly the same arguments as those of \cite{dempster:laird:rubin:1977} for the classic EM algorithm, it is straightforward to check that under suitable regularity assumptions, $T_{\Theta}$ is such that
\begin{enumerate}
\item The Kullback-Leibler divergence $D(\pi|f_{\theta}) = \int \log\frac{\pi(x)}{f_{\theta}(x)} \pi(x) \ud x$ is a Lyapunov function for the mapping $T_{\Theta}$, that is,
\[
  D\left(\pi|f_{T_{\Theta}(\theta)}\right) \leq D\left(\pi|f_{\theta}\right) \, .
\]
\item The set of fixed points of $T_{\Theta}$, i.e., such that $T_{\Theta}(\theta) = \theta$, is given by
\[
  \{\theta : \nabla_\theta D(\pi|f_{\theta}) = 0\} \, ,
\]
where $\nabla_\theta$ denotes the gradient.
\end{enumerate}
Obviously, \eqref{eq:limitingEM_param} is not directly exploitable in a statistical context as it
involves integrating under the unknown distribution of the observations. This limiting EM recursion
can however be used in the context of adaptive Monte Carlo methods
\citep{cappe:douc:guillin:marin:robert:2008} and is known in machine learning as part of the
\emph{information bottleneck} framework \citep{slonim_weiss_03-bottleneck}.

\subsection{Limitations of Batch EM for Long Data Records}
\label{sec:batchEM:limit}
But the main interest of \eqref{eq:limitingEM_param} is to provide a clear understanding of the
behaviour of the EM algorithm when used with very long data records, justifying much of the
intuition of \cite{neal:hinton:1999}. Note that a large part of the post 1980's literature on the EM
algorithm focusses on accelerating convergence towards the MLE for a fixed
data size. Here, we consider the related but very different issue of understanding the behaviour of
the EM algorithm when the data size increases.

\begin{figure}[hbt] \centering
  \includegraphics[width=0.7\textwidth]{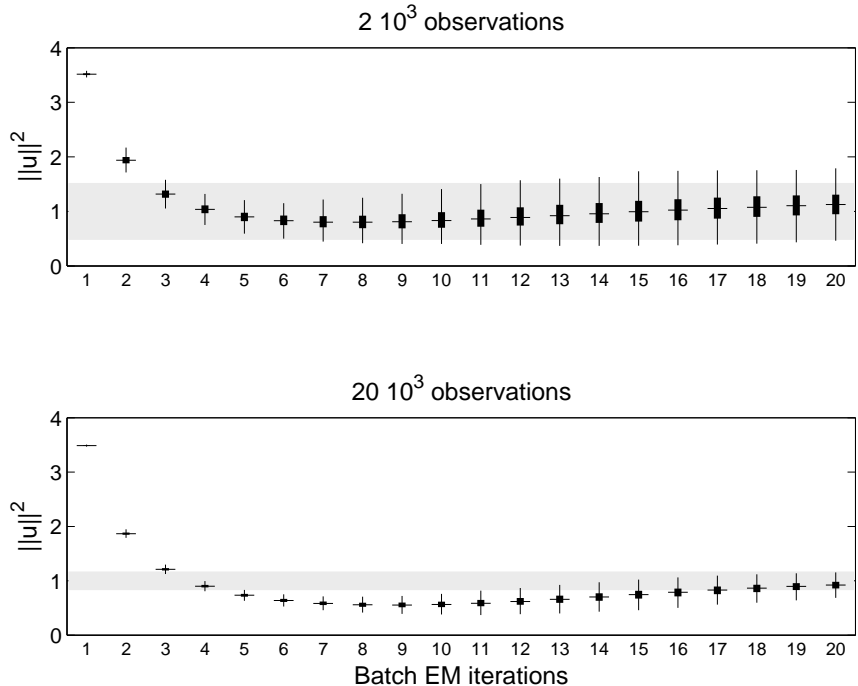}
  \caption{Convergence of batch EM estimates of $\|u\|^2$ as a function of the number of EM iterations for 2,000 (top) and 20,000 (bottom) observations. The box-and-whisker plots (outliers plotting suppressed) are computed from 1,000 independent replications of the simulated data. The grey region corresponds to $\pm 2$ interquartile range (approx. $99.3$\% coverage) under the asymptotic Gaussian approximation of the MLE.}
  \label{fig:pca1_batchEM}
\end{figure}

Figure~\ref{fig:pca1_batchEM} displays the results obtained with the batch EM algorithm for the
single component PCA model estimated from data simulated under the model with $u$ being a
20-dimensional vector of unit norm and $\lambda=5$. Both $u$ and $\lambda$ are treated as unknown
parameters but only the estimated squared norm of $u$ is displayed on
Figure~\ref{fig:pca1_batchEM}, as a function of the number of EM iterations and for two different
data sizes. Note that in this very simple model, the observation likelihood being given the
$\gaus(0,uu' + \lambda I_d)$ density, it is straightforward to check that the Fisher information
for the parameter $\|u\|^2$ equals $\{2 (\lambda + \|u\|^2)^2\}^{-1}$, which has been used to
represent the asymptotic confidence interval in grey. The width of this interval is meant to be
directly comparable to that of the whiskers in the box-and-whisker plots. Boxplots are used to
summarise the results obtained from one thousand independent runs of the method.

Comparing the top and bottom plots clearly shows that when the number of iterations increases, the
trajectories of the EM algorithm converge to a fixed deterministic trajectory which is that of the
limiting EM recursion $(T^k_{\Theta}(\theta_0))_{k\geq 1}$. Of course, this trajectory depends on
the choice of the initial point $\theta_0$ which was fixed throughout this experiment. It is also
observed that the monotone increase in the likelihood guaranteed by the EM algorithm does not
necessarily imply a monotone convergence of all parameters towards the MLE.  Hence, if the number
of EM iterations is kept fixed, the estimates returned by the batch EM algorithm with the larger
number of observations (bottom plot) are not significantly improved despite the ten-fold increase
in computation time (the E step computations must be done for all observations and hence the
complexity of the E-step scales proportionally to the number $n$ of observations). From a
statistical perspective, the situation is even less satisfying as estimation errors that were not
statistically significant for 2,000 observations can become significant when the number of
observations increases. In the upper plot, interrupting the EM algorithm after 3 or 4 iterations
does produce estimates that are statistically acceptable (comparable to the exact MLE) but in the
lower plot, about 20 iterations are needed to achieve the desired precision. As suggested by
\cite{neal:hinton:1999}, this paradoxical situation can to some extent be avoided by updating the
parameter (that is, applying the M-step) more often, without waiting for a complete scan of the
data record.

\section{Online Expectation-Maximisation}
\label{sec:onlineEM}
\subsection{The Algorithm}
The limiting-EM argument developed in the following section shows that when the number of observations tends to infinity, the EM algorithm is trying to locate the fixed points of the mapping $T_{\mathcal{S}}$ defined in~(\ref{eq:limitingEM}), that is, the roots of the equation
\begin{equation}
  \label{eq:onlineEM_onjective_S}
  \E_\pi \left(\E_{\bar{\theta}(S)} \left[\left. s(X_1,Y_1) \right| Y_1\right] \right) - S = 0 \, .  
\end{equation}
Although we cannot compute the required expectation wrt. the unknown distribution $\pi$ of the
data, each new observation provides us with an unbiased noisy observation of this quantity through
$\E_{\bar{\theta}(S)} \left[\left. s(X_n,Y_n) \right|
  Y_n\right]$. Solving~\eqref{eq:onlineEM_onjective_S} is thus an instance of the most basic case
where the Stochastic Approximation (or Robbins-Monro) method can be used. The literature on
Stochastic Approximation is huge but we recommend the textbooks by
\cite{benveniste:metivier:priouret:1990,kushner:yin:2003} for more details and examples as well as
the review paper by \cite{lai:2003} for a historical perspective. The standard stochastic approximation approach to approximate the solution of~\eqref{eq:onlineEM_onjective_S} is to compute
\begin{equation}
  \label{eq:tmp}
  S_n = S_{n-1} + \gamma_n \left(\E_{\bar{\theta}(S_{n-1})} \left[\left. s(X_n,Y_n) \right|
      Y_n\right] - S_{n-1}\right) \, ,  
\end{equation}
for $n \geq 1$, $S_0$ being arbitrary and $(\gamma_n)_{n\geq 1}$ denoting a sequence of positive \emph{stepsizes} that decrease to zero. This equation, rewritten in an equivalent form below, is the main ingredient of the online EM algorithm.

\begin{algorithm}[Online EM Algorithm] Given $S_0, \theta_0$ and a sequence of stepsizes $(\gamma_n)_{n\geq 1}$, do, for $n \geq 1$,
\label{alg:onlineEM}
\begin{description}
\item[Stochastic E-Step]
  \begin{equation}
    \label{eq:online:E}
  S_n = (1-\gamma_n) S_{n-1} + \gamma_n \E_{\theta_{n-1}} \left[\left. s(X_n,Y_n) \right|
    Y_n\right] \, ,    
  \end{equation}
\item[M-Step]
  \begin{equation}
    \label{eq:online:M}
  \theta_n = \bar{\theta}(S_n) \, .    
  \end{equation}
\end{description}
\end{algorithm}

Rewriting~\eqref{eq:onlineEM_onjective_S} under the form
displayed in~\eqref{eq:online:E} is very enlightening as it shows that the new statistic $S_n$ is
obtained as a convex combination of the previous statistic $S_{n-1}$ and of an update that depends
on the new observation $Y_n$. In particular it shows that the stepsizes
$\gamma_n$ have a natural scale as their highest admissible value is equal to one. This means that
one can safely take $\gamma_1=1$ and that only the rate at which the stepsize decreases needs to be
selected carefully (see below). It is also observed that if $\gamma_1$ is set to one, the initial
value of $S_0$ is never used and it suffices to select the \emph{initial parameter guess}
$\theta_0$; this is the approach used in the following simulations.

The only adjustment to Algorithm~\ref{alg:onlineEM} that is necessary in practice is to omit the
M-step of~(\ref{eq:online:M}) for the first observations. It typically takes a few observations
before the complete-data maximum likelihood solution is well-defined and the parameter update
should be inhibited during this early phase of training \citep{cappe:moulines:2009}. In the
simulations presented below in Sections~\ref{sec:prop} and~\ref{sec:batch_ML}, the M-step was
omitted for the first five observations only but in most complex scenarios a longer initial
parameter freezing phase may be necessary.

When $\gamma_1 < 1$, it is tempting to interpret the value of $S_0$ as being associated to a
prior on the parameters. Indeed, the choice of a conjugate prior for $\theta$ in the exponential
family defined by~(\ref{eq:complete_data_exp_family}) does result in a complete-data Maximum A
Posteriori (MAP) estimate of $\bar{\theta}$ given by $\bar{\theta}(S_{n,k}+S_0/n)$ (instead of
$\bar{\theta}(S_{n,k})$ for the MLE), where $S_0$ is the hyper-parameter of the prior
\citep{robert:2001}. However, it is easily checked that the influence of $S_0$ in $S_n$ decreases
as $\prod_{k=1}^n(1-\gamma_k)$, which for the suitable stepsize decrease schemes (see
beginning of Section~\ref{sec:prop} below), decreases faster than $1/n$. Hence, the value of
$S_0$ has a rather limited impact on the convergence of the online EM algorithm. To achieve MAP
estimation (assuming a conjugate prior on $\theta$) it is thus recommend to
replace~\eqref{eq:online:M} by $\theta_n = \bar{\theta}(S_n+S_0/n)$, where $S_0$ is the
hyperparameter of the prior on $\theta$.

A last remark of importance is that Algorithm~\ref{alg:onlineEM} can most naturally be interpreted
as a stochastic approximation recursion on the sufficient statistics rather than on the
parameters. There does not exists a similar algorithm that operates directly on the parameters
because unbiased approximations of the rhs. of (\ref{eq:limitingEM_param}) based on the
observations $Y_t$ are not easily available. As we will see below, Algorithm~\ref{alg:onlineEM} is
asymptotically equivalent to a gradient recursion on $\theta_n$ that involves an additional matrix
weighting which is not necessary in~\eqref{eq:online:E}.

\subsection{Convergence Properties}
\label{sec:prop}
Under the assumption that the stepsize sequence satisfies $\sum_n \gamma_n = \infty$, $\sum_n
\gamma_n^2 < \infty$ and other regularity hypotheses that are omitted here (see
\citealp{cappe:moulines:2009} for details) the following properties characterise the asymptotic behaviour of the online EM algorithm.
\begin{enumerate}
\item[(i)] The estimates $\theta_n$ converge to the set of roots of the equation $\nabla_\theta D(\pi|f_{\theta}) = 0$.
\item[(ii)] The algorithm is asymptotically equivalent to a gradient algorithm
  \begin{equation}
  \label{eq:gradalg}
  \theta_{n} = \theta_{n-1} + \gamma_{n} J^{-1}(\theta_{n-1}) \nabla_\theta\log
  f_{\theta_{n-1}}(Y_{n}) \, ,
  \end{equation}
  where $J(\theta) = - \E_\pi\left( \E_\theta \left[ \nabla_\theta^2 \log p_\theta(X_1,Y_1) \big| Y_1 \right] \right)$.
\item[(iii)] For a well specified model (i.e., if $\pi = f_{\theta_\star}$) and under
  \emph{Polyak-Ruppert averaging}, $\theta_n$ is Fisher efficient: sequences that do converge to
  $\theta_\star$ are such that $\sqrt{n}(\theta_n-\theta_\star)$ converges in distribution to a
  centred multivariate Gaussian variable with covariance matrix $I_\pi(\theta_\star)$, where
  $I_\pi(\theta) = -\E_\pi[\nabla^2_\theta \log f_{\theta}(Y_1)]$ is the Fisher information
  matrix corresponding to the observed data.
\end{enumerate}

Polyak-Ruppert averaging refers to a postprocessing step which simply consists in replacing the estimated parameter values $\theta_{n}$ produced by the algorithm by their average
\[
  \tilde{\theta}_n = \frac1{n-n_0} \sum_{t=n_0+1}^n \theta_n \, ,
\]
where $n_0$ is a positive index at which averaging is started
\citep{polyak:juditsky:1992,ruppert:1988}. Regarding the statements (i) and
(ii) above, it is important to understand that the limiting estimating equation
$\nabla_\theta D(\pi|f_{\theta}) = 0$ may have multiple solutions, even in
well-specified models. In practice, the most important factor that influences
the convergence to one of the stationary points of the Kullback-Leibler
divergence $D(\pi|f_{\theta})$ rather than the other is the choice of the
initial value $\theta_0$. An additional important remark about (i)--(iii) is
the fact that the asymptotic equivalent gradient algorithm
in~\eqref{eq:gradalg} is not a practical algorithm as the matrix $J(\theta)$
depends on $\pi$ and hence cannot be computed. Note also that $J(\theta)$ is
(in general) neither equal to the complete-data information matrix
$I_p(\theta)$ nor to the actual Fisher information in the observed model
$I_\pi(\theta)$. The form of $J(\theta)$ as well as its role to approximate the
convergence behaviour of the EM algorithm follows the idea of
\cite{lange:1995}.

\begin{figure}[hbt] \centering
  \includegraphics[width=0.7\textwidth]{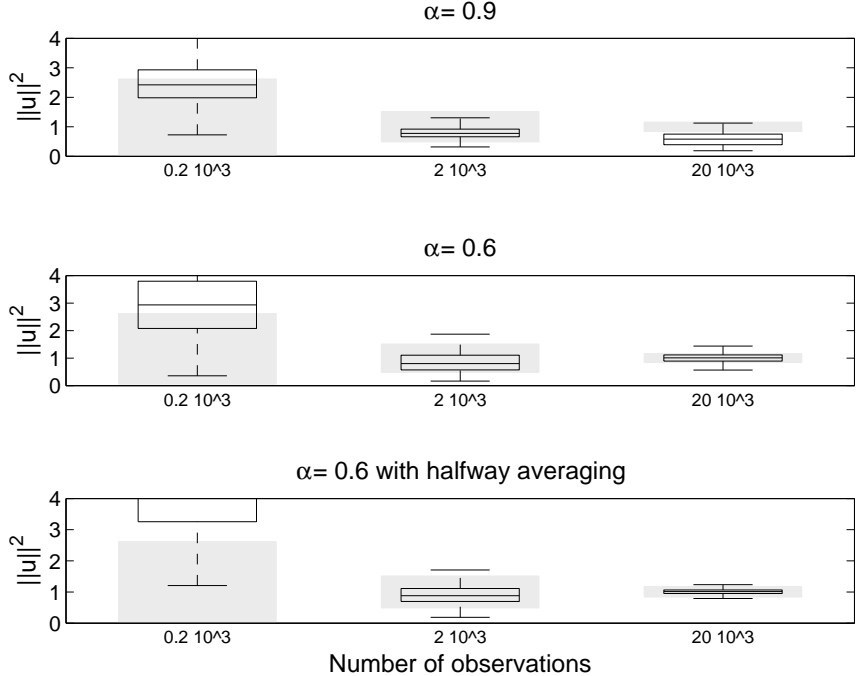}
  \caption{Online EM estimates of $\|u\|^2$ for various data sizes (200, 2,000 and 20,000 observations, from left to right) and algorithm settings ($\alpha=0.9$, $\alpha=0.6$ and $\alpha=0.6$ with Polyak-Ruppert averaging, from top to bottom). The box-and-whisker plots (outliers plotting suppressed) are computed from 1,000 independent replications of the simulated data. The grey regions corresponds to $\pm 2$ interquartile range (approx. $99.3$\% coverage) under the asymptotic Gaussian approximation of the MLE.}
  \label{fig:pca1_online}
\end{figure}

From our experience, it is generally sufficient to consider stepsize sequences of the form
$\gamma_n = 1/n^{\alpha}$ where the useful range of values for $\alpha$ is from 0.6 to
0.9. The most robust setting is obtained when taking $\alpha$ close to $0.5$ and using Polyak-Ruppert
averaging. The latter however requires to chose an index $n_0$ that is sufficiently large and,
hence, some idea of the convergence time is necessary. To illustrate these observations,
Figure~\ref{fig:pca1_online} displays the results of online EM estimation for the single component
PCA model, exactly in the same conditions as those considered previously for batch EM estimation in
Section~\ref{sec:batchEM:limit}.

From a computational point of view, the main difference between the online EM algorithm and the
batch EM algorithm of Section~\ref{sec:batch_EM} is that the online algorithm performs the M-step
update in~\eqref{eq:pca1_M} after each observation, according to~\eqref{eq:online:M}, while the
batch EM algorithm only applies the M-step update after a complete scan of all available
observations. Both algorithms however require the computation of the E-step statistics
following~\eqref{eq:pca1_E} for each observation. In batch EM, these local E-step computation are
accumulated, following~\eqref{eq:batchEM_E}, while the online algorithm recursively averages these
according to~(\ref{eq:online:E}). Hence, as the computational complexity of the E and M steps are,
in this case, comparable, the computational complexity of the online estimation is equivalent to
that of one or two batch EM iterations. With this in mind, it is obvious that the results of
Figure~\ref{fig:pca1_online} compare very favourably to those of Figure~\ref{fig:pca1_batchEM} with
an estimation performance that is compatible with the statistical uncertainty for observation
lengths of 2,000 and larger (last two boxplots on the right in each display).

Regarding the choice of the stepsize decrease exponent $\alpha$, it is observed that while the
choice of $\alpha=0.6$ (middle plot) does result in more variability than the choice of
$\alpha=0.9$ (top plot), especially for smaller observation sizes, the long-run performance is
somewhat better with the former. In particular, when Polyak-Ruppert averaging is used (bottom
plot), the performance for the longest data size (20,000 observations) is clearly compatible with
the claim that online EM is Fisher-efficient in this case. A practical concern associated with
averaging is the choice of the initial instant $n_0$ where averaging starts. In the case of
Figure~\ref{fig:pca1_online}, we choose $n_0$ to be equal to half the length of each data record,
hence averaging is used only on the second half of the data. While it produces refined estimates
for the longer data sizes, one can observe that the performance is rather degraded for the smallest
observation size (200 observations) due to the fact that the algorithm is still very far from
having converged after just 100 observations. Hence, averaging is efficient but does require to
chose a value of $n_0$ that is sufficiently large so as to avoid introducing a bias due
to the lack of convergence.

\begin{figure}[hbtp] \centering
  \includegraphics[width=0.7\textwidth]{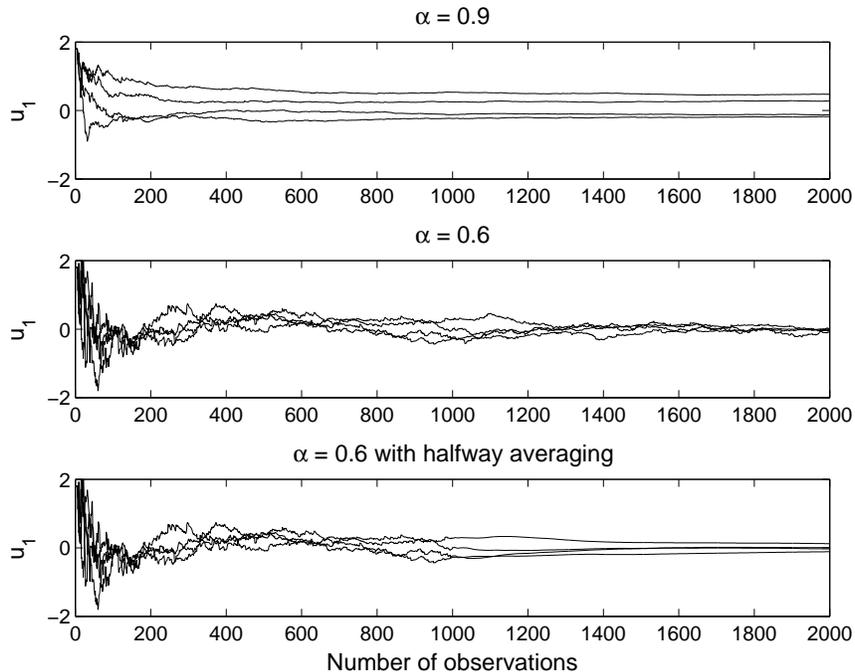}
  \caption{Four superimposed trajectories of the estimate of $u_1$ (first component of $u$) for various algorithm settings ($\alpha=0.9$, $\alpha=0.6$ and $\alpha=0.6$ with Polyak-Ruppert averaging, from top to bottom). The actual value of $u_1$ is equal to zero.}
  \label{fig:pca1_traj}
\end{figure}

In our experience, the fact that choices of $\alpha$ close to 0.5 are more reliable than values
closer to the upper limit of 1 is a very constant observation. It may come to some surprise for
readers familiar with the gradient descent algorithm as used in numerical optimisation, which
shares some similarity with~\eqref{eq:online:E}. In this case, it is known that the optimal stepsize
choice is of the form $\gamma_n = a (n + b)^{-1}$ for a broad class of functions
\citep{nesterov:2003}. However, the situation here is very different as we do not observe exact
gradient or expectations but only noisy version of them\footnote{For a complete-data exponential
  family model in natural parameterisation, it is easily checked that $\E_{\theta}
  \left[\left. s(X_n,Y_n) \right| Y_n\right] = \E_{\theta}[s(X_n,Y_n)] + \nabla_\theta \log
  f_\theta (Y_n)$ and hence that the recursion in the space of sufficient statistics is indeed very
  close to a gradient ascent algorithm. However, we only have access to
  $\nabla_\theta \log f_\theta (Y_n)$ which is a noisy version of the gradient of the actual limiting objective function
  $D(\pi|f_{\theta})$ that is minimised.}. Figure~\ref{fig:pca1_traj} shows that
while the trajectory of the parameter estimates appears to be much rougher and variable with
$\alpha=0.6$ than with $\alpha=0.9$, the bias caused by the initialisation is also forgotten much
more rapidly. It is also observed that the use of averaging (bottom display) makes it possible to
achieve the best of both worlds (rapid forgetting of the initial condition and smooth
trajectories).

\cite{liang:klein:2009} considered the performance of the online EM algorithm for large-scale
natural language processing applications. This domain of application is characterised by the use of
very large-dimensional models, most often related to the multinomial distribution, involving tens
of thousands of different words and tens to hundreds of semantic tags. As a consequence, each
observation, be it a sentence or a whole document, is poorly informative about the model parameters
(typically a given text contains only a very limited portion of the whole available vocabulary). In
this context, \cite{liang:klein:2009} found that the algorithm was highly competitive with other
approaches but only when combined with \emph{mini-batch blocking}: rather than applying
Algorithm~\ref{alg:onlineEM} at the observation scale, the algorithm is used on mini-batches 
consisting of $m$ consecutive observations $(Y_{m (k-1)+1}, Y_{m k+2} \dots, Y_{m k})_{k \geq
  1}$. For the models and data considered by \cite{liang:klein:2009}, values of $m$ of up to a few
thousands yielded optimal performance. More generally, mini-batch blocking can be useful in dealing
with mixture-like models with rarely active components.

\subsection{Application to Finite Mixtures}
Although we considered so far only the simple case of Example~\ref{ex:pca1} which allows for the
computation of the Fisher information matrix and hence for quantitative assessment of the asymptotic
performance, the online EM algorithm is easy to implement in models involving finite mixture of
distributions.

\begin{figure}[t] \centering
  \includegraphics[width=0.3\textwidth]{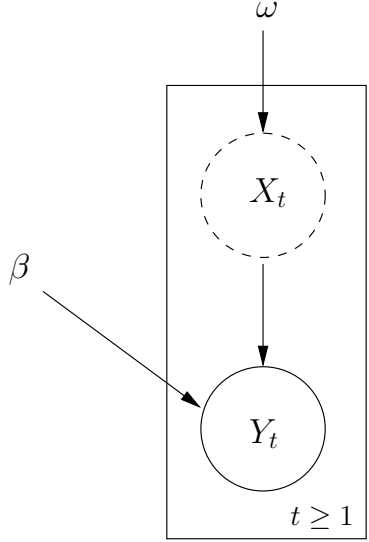}
  \caption{Bayesian network representation of a mixture model.}
  \label{fig:mix-bn}
\end{figure}

Figure~\ref{fig:mix-bn} displays the Bayesian network representation corresponding to a
mixture model: for each observation $Y_t$ there is an unobservable mixture indicator or allocation
variable $X_t$ that takes its value in the set $\{1,\dots,m\}$. A typical parameterisation for this
model is to have a separate sets of parameters $\omega$ and $\beta$ for, respectively, the
parameters of the prior on $X_t$ and of the conditional distribution of $Y_t$ given $X_t$. Usually,
$\omega = (\omega^{(1)}, \dots, \omega^{(m)})$ is chosen to be the collection of component
frequencies, $\omega^{(i)} = \P_\theta(X_t = i)$, and hence $\omega$ is constrained to the
probability simplex ($\omega^{(i)} \geq 0$ and $\sum_{i=1}^m \omega^{(i)} = 1$). The observation
pdf is most often parameterised as
\[
  f_\theta(y_t) = \sum_{i=1}^m \omega^{(i)} g_{\beta^{(i)}}(y_t) \, ,
\]
where $(g_\lambda)_{\lambda \in \Lambda}$ is a parametric family of probability densities and
$\beta = (\beta^{(1)}, \dots, \beta^{(m)})$ are the component-specific parameters. We assume that
$g_\lambda(y_t)$ has an exponential family representation similar to that
of~(\ref{eq:complete_data_exp_family}) with sufficient statistic $s(y_t)$ and maximum likelihood
function $\bar{\lambda} : \mathcal{S} \mapsto \Lambda$, which is such that
$\bar{\lambda}(\frac{1}{n}\sum_{t=1}^n s(Y_t)) = \arg\max_{\lambda \in \Lambda}
\frac{1}{n}\sum_{t=1}^n \log g_\lambda(Y_t)$. It is then easily checked that the complete-data
likelihood $p_\theta$ belongs to an exponential family with sufficient statistics
\begin{align*}
  & s^{(\omega,i)}(X_t) = \1\{X_t = i\} \, , \\
  & s^{(\beta,i)}(X_t,Y_t) = \1\{X_t = i\} s(Y_t)\, , \quad \text{for $i=1,\dots,m$.}
\end{align*}
And the function $\bar{\theta}(S)$ can then be decomposed as
\begin{align*}
  & \omega^{(i)} = S^{(\omega,i)}\, , \\
  & \beta^{(i)} = \bar{\lambda}\left(S^{(\lambda,i)}/S^{(\omega,i)}\right) \, , \quad \text{for $i=1,\dots,m$.}.
\end{align*}
Hence the online EM algorithm takes the following specific form.

\begin{algorithm}[Online EM Algorithm for Finite Mixtures] Given $S_0, \theta_0$ and a sequence of stepsizes $(\gamma_n)_{n\geq 1}$, do, for $n \geq 1$,
\label{alg:onlineEM_mix}
\begin{description}
\item[Stochastic E-Step] Compute
  \begin{equation*}
    P_{\theta_{n-1}}(X_t = i|Y_t) = \frac{\omega_{n-1}^{(i)} \, g_{\beta_{n-1}^{(i)}}(Y_t)}{\sum_{j=1}^m \omega_{n-1}^{(j)} \, g_{\beta_{n-1}^{(j)}}(Y_t)} \, ,
  \end{equation*}
  and
  \begin{align}
    & S_n^{(\omega,i)} = (1-\gamma_n)S_{n-1}^{(\omega,i)} + \gamma_n P_{\theta_{n-1}}(X_t = i|Y_t) \, , \nonumber \\
    & S_n^{(\beta,i)} =  (1-\gamma_n)S_{n-1}^{(\beta,i)} + \gamma_ns(Y_t) P_{\theta_{n-1}}(X_t = i|Y_t) \, , \label{eq:online_mix_E}
  \end{align}
  for $i=1,\dots, m$.
\item[M-Step]
  \begin{align}
    & \omega_{n}^{(i)} = S_n^{(\omega,i)} \, , \nonumber \\
    & \beta_{n}^{(i)} = \bar{\lambda}\left(S^{(\lambda,i)}_n/S^{(\omega,i)}_n\right) \, . \label{eq:online_mix_M}
  \end{align}
\end{description}
\end{algorithm}

\begin{example}[Online EM for Mixtures of Poisson]
  \label{ex:poisson_mix}
  We consider a simplistic instance of Algorithm~\ref{alg:onlineEM_mix} corresponding to the
  mixture of Poisson distribution (see also Section 2.4 of \citealp{cappe:moulines:2009}). In the
  case of the Poisson distribution, $g_\lambda(y_t) = \frac1{y_t!} \ue^{-\lambda} \lambda^y_t$, the
  sufficient statistic reduces to $s(y_t) = y_t$ and the MLE function $\bar{\lambda}$ also is the
  identity $\bar{\lambda}(S) = S$. Hence, the online EM recursion for this case simply consists in
  instantiating \eqref{eq:online_mix_E}--\eqref{eq:online_mix_M} as
  \begin{align*}
    & S_n^{(\omega,i)} = (1-\gamma_n)S_{n-1}^{(\omega,i)} + \gamma_n P_{\theta_{n-1}}(X_t = i|Y_t) \, , \\
    & S_n^{(\beta,i)} = (1-\gamma_n)S_{n-1}^{(\beta,i)} + \gamma_n Y_t P_{\theta_{n-1}}(X_t = i|Y_t) \, ,
  \end{align*}
  and
  \begin{align*}
    & \omega_{n}^{(i)} = S_n^{(\omega,i)} \, , \\
    & \beta_{n}^{(i)} = S^{(\lambda,i)}_n/S^{(\omega,i)}_n \, .
  \end{align*}
\end{example}

\subsection{Use for Batch Maximum-Likelihood Estimation}
\label{sec:batch_ML}
An interesting issue that deserves some more comments is the use of the online EM algorithm for
batch estimation from a fixed data record $Y_1,\dots, Y_N$. In this case, the objective is to save
on computational effort compared to the use of the batch EM algorithm.

The analysis of the convergence behaviour of online EM in this context is made easy by
the following observation: Properties (i) and (ii) stated at the beginning of
Section~\ref{sec:prop} do not rely on the assumption that the model is well-specified (i.e., that
$\pi = f_{\theta_\star}$) and can thus be applied with $\pi$ being the empirical distribution
$\hat{\pi}_N(dy) = \frac1N \sum_{t=1}^N \delta_{Y_t}(dy)$ associated to the observed
sample\footnote{Where $\delta_{u}(dy)$ denotes the Dirac measure localised in $u$.}. Hence, if
the online EM algorithm is applied by randomly drawing (with replacement) subsequent
``pseudo-observations'' from the finite set $\{Y_1,\dots,Y_N\}$, it converges to points such that
\[
  \nabla_\theta D(\hat{\pi}_N|f_{\theta}) = -\frac1N \nabla_\theta \sum_ {t=1}^N \log f_{\theta}(Y_t)
=0 \, ,
\]
that is, stationary points of the log-likelihood of the observations $Y_1, \dots,
Y_N$. Property (ii) also provides an asymptotic equivalent of the online EM update, where the index
$n$ in~(\ref{eq:gradalg}) should be understood as the number of online EM steps rather than the
number of actual observation, which is here fixed to $N$.

In practice, it doesn't appear that drawing randomly the pseudo-observations make any real
difference when the observations $Y_1,\dots,Y_N$ are themselves already independent, except for
very short data records. Hence, it is more convenient to scan the observation systematically in
\emph{tours} of length $N$ in which each observation is visited in a predetermined order. At the
end of each tour, $k = n/N$ is equal to the number of batch tours completed since the start of the
online EM algorithm.

To compare the numerical efficiency of this approach with that of the usual batch EM it is
interesting to bring together two results. For online EM, based on~(\ref{eq:gradalg}), it is
possible to show that $\sqrt{n} (\theta_n - \theta_\infty)$ converges in distribution to a
multivariate Gaussian distribution, where $\theta_\infty$ denotes the limit of
$\theta_n$\footnote{Strictly speaking, this has been shown only for random batch scans.}. In
contrast, the batch EM algorithm achieves so-called linear convergence which means that for
sufficiently large $k$'s, there exists $\rho < 1$ such that $\| \theta_k - \theta_\infty \| \leq
\rho^k$, where $\theta_k$ denotes the parameter estimated after $k$ batch EM iterations
\citep{dempster:laird:rubin:1977,lange:1995}. In terms of computing effort, the number $k$ of batch
EM iterations is comparable to the number $k = n/N$ of batch tours in the online EM
algorithm. Hence the previous theoretical results suggest that
\begin{itemize}
\item If the number of available observations $N$ is small, batch EM can be way faster than the
  online EM algorithm, especially if one wants to obtain a very accurate numerical approximation of
  the MLE. Note that from a statistical viewpoint, this may be unnecessary as the MLE itself is
  only a proxy for the actual parameter value, with an error that is of order $1/\sqrt{N}$ in
  regular statistical models.
\item When $N$ increases, the online EM algorithm becomes preferable and, indeed, arbitrary so if
  $N$ is sufficient large. Recall in particular from Section~\ref{sec:prop} that when $N$
  increases, the online EM estimate obtained after a \emph{single batch tour} is asymptotically
  equivalent to the MLE whereas the estimate obtained after a single batch EM iteration converges
  to the deterministic limit $T_{\Theta}(\theta_0)$.
\end{itemize}

In their pioneering work on this topic, \cite{neal:hinton:1999} suggested an algorithm
called \emph{incremental EM} as an alternative to the batch EM algorithm. The incremental EM
algorithm turns out to be exactly equivalent to Algorithm~\ref{alg:onlineEM} used with $\gamma_n =
1/n$ up to the end of the batch tour only. After this initial batch scan, the incremental EM
proceeds a bit differently by replacing one by one the previously computed values of
$\E_{\theta_{t-1}}\left[\left. s(X_t,Y_t) \right| Y_t\right]$ with
$\E_{\theta_{t-1}+(k-1)N}\left[\left. s(X_t,Y_t) \right| Y_t\right]$ when processing the
observation at position $t$ for the $k$-th time. This incremental EM algorithm is indeed more
efficient than batch EM, although they do not necessarily have the same complexity ---a point that
will be further discussed in the next section. For large values of $N$ however, incremental EM
becomes impractical (due to the use of a storage space that increases proportionally to $N$) and
less recommendable than the online EM algorithm as shown by the following example (see also the
experiments of \citealp{liang:klein:2009} for similar conclusions).

\begin{figure}[hbtp] \centering
  \includegraphics[width=0.7\textwidth]{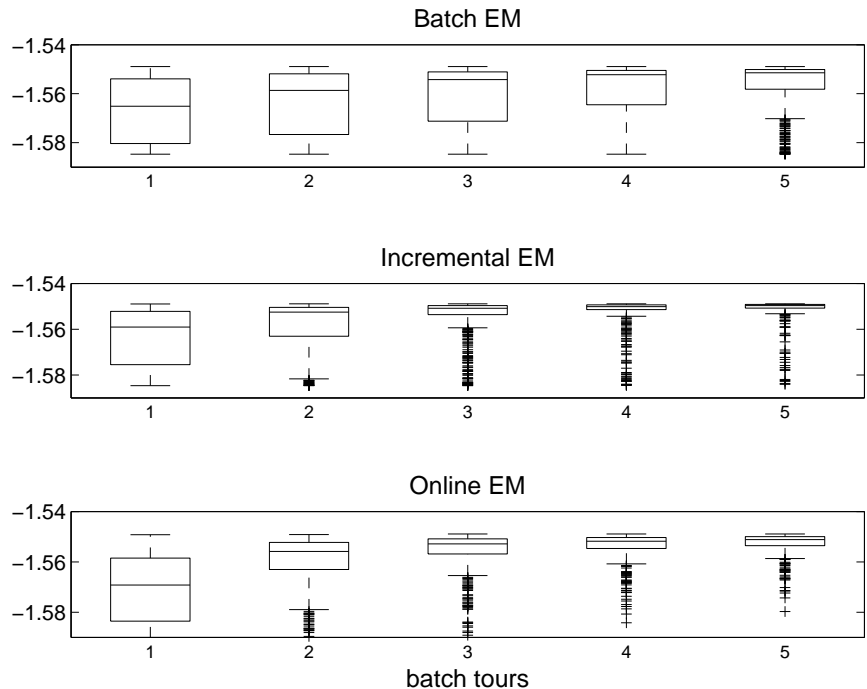}
  \caption{Normalised log-likelihood of the estimates obtained with, from top to bottom, batch EM, incremental EM and online EM as a function of the number of batch tours (or iterations, for batch EM). The data is of length $N=100$ and the box an whiskers plots summarise the results of 500 independent runs of the algorithms started from randomised starting points $\theta_0$.}
  \label{fig:pm_comp_100}
\end{figure}

\begin{figure}[hbtp] \centering
  \includegraphics[width=0.7\textwidth]{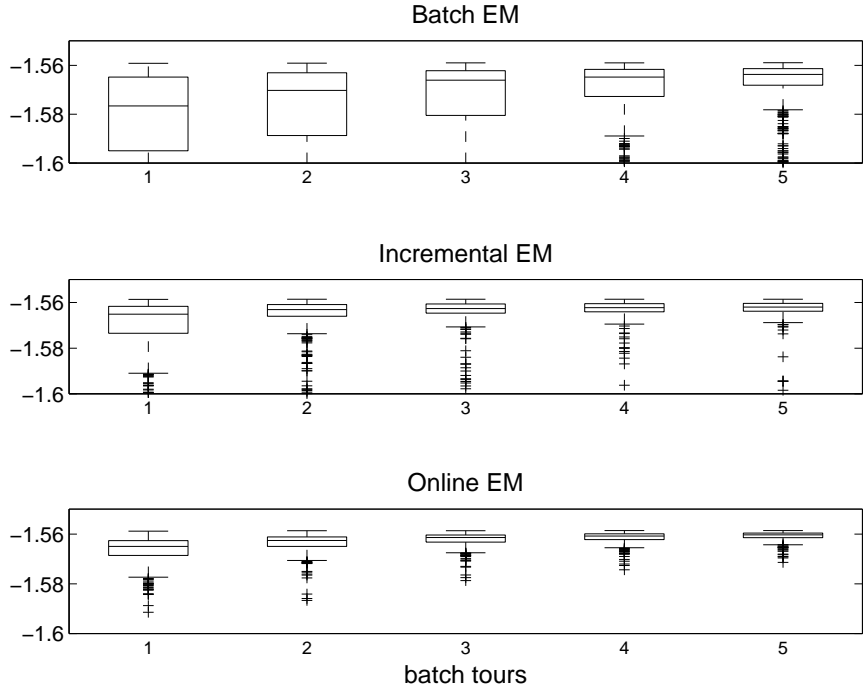}
  \caption{Same display as in Figure~\ref{fig:pm_comp_100} for a data record of length $N=\text{1,000}$.}
  \label{fig:pm_comp_1000}
\end{figure}

Figure~\ref{fig:pm_comp_100} displays the normalised log-likelihood values corresponding to the
three estimation algorithms (batch EM, incremental EM and online EM) used to estimate the
parameters of a mixture of two Poisson distributions (see Example~\ref{ex:poisson_mix} for
implementation details)\footnote{Obviously, the fact that only log-likelihoods normalised by the
  length of the data are plotted hides some important aspects of the problem, in particular the
  lack of identifiability caused by the unknown labelling of the mixture components.}. All data is
simulated from a mixture model with parameters $\omega^{(1)}=0.8$, $\beta^{(1)}=1$ and
$\beta^{(2)}=3$. In this setting, where the sample size is fixed, it is more difficult to come up
with a faithful illustration of the merits of each approach as the convergence behaviour of the
algorithms depend very much on the data record and on the initialisation. In
Figures~\ref{fig:pm_comp_100} and~\ref{fig:pm_comp_1000}, the two data sets were kept fixed throughout the simulations but the initialisation of both Poisson means was randomly chosen from the
interval [0.5,5]. This randomisation avoids focussing on particular algorithm trajectories and
gives a good idea of the general situation, although some variations can still be observed when
varying the observation records.

Figure~\ref{fig:pm_comp_100} corresponds to the case of an observation sequence of length
$N=100$. In this case it is observed that the performance of incremental EM dominates that of the
other two algorithms, while the online EM algorithm only becomes preferable to batch EM after the
second batch tour. For the data record of length $N=\text{1,000}$ (Figure~\ref{fig:pm_comp_1000}), online
EM now dominates the other two algorithms, with incremental still being preferable to batch
EM. Notice that it is also the case after the first batch tour, illustrating our claim that the
choice of $\gamma_n=n^{-0.6}$ used here for the online EM algorithm is indeed preferable to the
value of $\gamma_n=n^{-1}$, which coincides with the update used by the incremental EM algorithm
during the first batch tour. Finally, one can observe on Figure~\ref{fig:pm_comp_1000} that even
after five EM iterations there are a few starting values for which batch EM or incremental EM gets
stuck in regions of low normalised log-likelihood (around -1.6). These indeed correspond to local
maxima of the log-likelihood and some trajectories of batch EM converge to these regions,
depending on the value of initialisation. The online EM with the above choice of step-size appears
to be less affected by this issue, although we have seen that in general only convergence to a
stationary point of the log-likelihood is guaranteed.

\section{Discussion}
\label{sec:disc}
We conclude this chapter by a discussion of the online EM
algorithm.

First, the approach presented here is not the only option for online
parameter estimation in latent variable models. One of the earliest and most
widely used (see, e.g., \citealp{liu:almhana:choulakian:mcgorman:2006})
algorithm is that of \cite{titterington:1984} consisting of the following gradient update:
\begin{equation}
  \label{eq:titt}
\theta_{n} = \theta_{n-1} + \gamma_{n} I^{-1}_p(\theta_{n-1}) \nabla_\theta\log
f_{\theta_{n-1}}(Y_{n}) \, ,  
\end{equation}
where the matrix $I_p$ refers to the complete-data Fisher information
matrix. For complete-data models in natural parameterisation ---with $\psi
\equiv 1$ in (\ref{eq:complete_data_exp_family}), $I_p(\theta)$ coincides with
$\nabla_\theta^2 A(\theta)$ and, as it does not depend on $X_t$ or $Y_t$, is
also equal to the matrix $J(\theta)$ that appears in~(\ref{eq:gradalg}). Thus,
Titterington's algorithm is in this case asymptotically equivalent to
online EM. In other cases,
where $I_p$ and $J$ differs, the recursion of~\eqref{eq:titt} converges at the
same rate as the online EM algorithm but is not Fisher-efficient. Another
difference is the way Algorithm~\ref{alg:onlineEM} deals with parameter
constraints. Assumption~\ref{asum:main} implies that the M-step update, taking
into account the possible parameter constraints, is explicit. Hence, $\theta_n
= \bar{\theta}(S_n)$ does satisfy the parameter constraint by definition of the
function $\bar{\theta}$. In the case of Example~\ref{ex:poisson_mix} for
instance, the proposed update does guarantee that the mixture weight vector
$\omega$ stays in the probability simplex. This is not the case with the update
of~\eqref{eq:titt} which requires reparameterisation or reprojection to handle
possible parameter constraints.

As discussed in Section~\ref{sec:batch_ML}, the online EM algorithm is inspired
by the work of \cite{neal:hinton:1999} but distinct from their incremental EM
approach. To the best of our knowledge, the online EM algorithm was first
proposed by~\cite{sato:2000} and~\cite{sato:ishii:2000} who described the algorithm and
provided some analysis of convergence in the case of exponential families in
natural parameterisation and for the case of mixtures of Gaussians.

In Section~\ref{sec:batch_ML}, we have seen that the online EM algorithm is
preferable to batch EM, in terms of computational effort, when the observation
size is sufficiently large. To operate in batch mode, the online EM needs to be
used repeatedly by scanning the data record several times so as to converge
toward the maximum likelihood estimate. It is important to note however that
one iteration of batch EM operating on $N$ observations requires $N$ individual
E-step computations but only one application of the M-step update; whereas
online EM applies the M-step update at each observation and, hence, requires
$N$ M-step updates per complete batch scan. Thus, the comparison between both
approaches also depends on the respective numerical complexities of the E and M
steps. The online approach is more attractive for models in which the M-step
update is relatively simple. The availability of parallel computing
resources would be most favourable to batch EM for which the E-step
computations pertaining to each observation may be done independently. In
contrast, in the online approach the computations are necessarily sequential
as the parameter is updated when processing each observation.

As indicated in the beginning of this chapter, the main strength of the online
EM algorithm is its simplicity and ease of implementation. As discussed above,
this is particularly true in the presence of constraints on the parameter
values. We have also seen in Section~\ref{sec:prop}, that
Algorithm~\ref{alg:onlineEM} is very robust with respect to the choice of the
stepsize $\gamma_n$. In particular, the absolute scale of the stepsize is fixed
due to the use of a convex combination in~(\ref{eq:online:E}) and it is
generally sufficient to consider stepsizes of the form $\gamma_n =
n^{-\alpha}$. We have shown that values of $\alpha$ of the order of 0.6 (i.e.,
closer to the lower limit of 0.5 than to the upper one of 1) yield more robust
convergence. In addition, Polyak-Ruppert averaging can be used to smooth the
parameter estimates and reach the optimal asymptotic rate of convergence that
makes the online estimate equivalent to the actual MLE.

As illustrated by Figure~\ref{fig:pca1_online}, the online EM algorithm is not
optimal for short data records of, say, less than 100 to 1,000 observations and
in this case performing batch mode estimation by repeatedly scanning the data
is recommended (see Figure~\ref{fig:pm_comp_100} for the typical improvement to
expect from this procedure). The main limitation of the online EM algorithm is
to require that the M-step update $\bar{\theta}$ be available in
closed-form. In particular, it would be interesting to extend the approach to
cases where $\bar{\theta}$ needs to be determined numerically, thus making it possible to handle
mixtures of Generalised Linear Models for instance (and not only mixtures of
linear regressions as in \citealp{cappe:moulines:2009}).

We finally mention two more directions in which recent works have proposed to
extend the online EM framework. The first one concerns the case of
non-independent observations and, in particular, of observations that follows a
Hidden Markov Model (HMM) with Markov dependencies between successive
states. \cite{mongillo:deneve:2008} and \cite{cappe:2009} have proposed an
algorithm for HMMs that appears to be very reminiscent of
Algorithm~\ref{alg:onlineEM}, although not directly interpretable as a
stochastic approximation recursion on the expected sufficient statistics. The
other topic of importance is motivated by the many cases where the E-step
computation of $\E_{\theta_{n-1}} \left[\left. s(X_n,Y_n) \right| Y_n\right]$
is not feasible. This typically occurs when the hidden variable $X_n$ is a
continuous variable. For such cases, a promising solution consists in
approximating the E-step using some form of Monte Carlo simulations (see, e.g.,
\citealp{cappe_09-onlinesmc-ssp,delmoral_doucet_singh_09} for methods that use
sequential Monte Carlo approaches).
However much conceptual and theoretical work remains to be done, as the consistency result
summarised in Section~\ref{sec:prop} only extends straightforwardly to the ---rather limited---
case of independent Monte Carlo draws $X_n^{(\theta_{n-1})}$ from the conditionals
$p_{\theta_{n-1}}(x_n|Y_n)$. In that case, $s(X_n^{(\theta_{n-1})},Y_n)$ still provides an unbiased
estimate of the limiting EM mapping in $\theta_{n-1}$ and the theory is very similar. In the
more general case where Markov chain or sequential Monte Carlo simulations are used to
produce the draws $X_n^{(\theta_{n-1})}$, the convergence of the online estimation procedure needs
to be investigated with care.

\bibliography{onlineem-arxiv}

\end{document}